\documentclass[12pt]{article}

\usepackage{graphicx} 
\usepackage{amsmath, amssymb} 

\usepackage{caption}       
\usepackage{setspace}
\usepackage{color,soul} 
\usepackage{hyperref} 
\usepackage{xcolor}
\usepackage[numbers,sort&compress]{natbib}
\usepackage[left]{lineno}
\usepackage[bottom=1.2in]{geometry}
\usepackage{float}

\newcommand{\de}{\mathrm{d}}

\newcommand{\ii}{\imath}

\newcommand{\kB}{k_\mathrm{B}}
\newcommand{\ve}{\varepsilon}

\begin{document}
\begin{center}
  {\LARGE Strong Epsilon-Near-Zero Confinement and Interface-Controlled Dynamics in Inverse-Designed Transparent Conducting Oxide Cavities}
  \vspace{1em}

  Jae Ik Choi$^{1,2}$, Ludmila J. Prokopeva$^{2}$, Vahagn Mkhitaryan$^{2}$, Yuheng Chen$^{2}$,\\
  Alexander V. Kildishev$^{2}$, Vladimir M. Shalaev$^{2}$, Alexandra Boltasseva$^{1,2*}$
  \vspace{1em}

    {\small
    $^1$School of Materials Engineering, Purdue University, West Lafayette, IN 47907, USA\\
    $^2$Elmore Family School of Electrical and Computer Engineering, Birck Nanotechnology Center, Purdue University, West Lafayette, IN 47907, USA\\ [0.5em]
    * aeb@purdue.edu\\
    }
\end{center}

\onehalfspacing

\begin{abstract}
Achieving optical modulation that is both strong and ultrafast remains a central challenge in time-varying photonics. Here, we theoretically demonstrate that combining inverse-designed cavities that utilize transparent conducting oxides (TCOs) operating at the epsilon-near-zero (ENZ) point with boundaries that support carrier transport can meet this requirement. This is achieved by leveraging ENZ-enhanced optical absorption and electron heating with interface-engineered carrier relaxation. A multilayer inverse-design framework, combined with a structural disorder-aware correction loop, is developed to preserve robust resonant absorption in ultrathin TCO films under realistic material disorder. Relative to conventional TCO cavity geometries, the inverse-designed structures exhibit stronger light confinement, greater electron temperature modulation, and markedly faster recovery. Beyond cavity design, interfacial barrier engineering and adjacent-layer doping emerge as effective degrees of freedom for co-designing spectral and temporal dynamics. Furthermore, our theoretical framework shows that the inverse-designed cavity supports richer oscillatory transmittance dynamics on the few-optical-cycle timescale compared to its single-film counterpart, with interfacial control enabling further temporal compression. The combined control of optical confinement and interfacial dynamics establishes inverse-designed TCO cavities as a viable platform for time-varying photonic media capable of supporting strong and ultrafast optical responses.
\end{abstract}


\section{Introduction}\label{sec1}
Achieving optical modulation that is both extremely fast and strong in magnitude is a defining challenge for emerging photonic technologies. Pushing the modulation timescale toward the optical-cycle regime fundamentally alters the nature of light--matter interactions, enabling non-adiabatic responses that are inaccessible to conventional photonic devices. Such extreme temporal control could enable time-domain manipulation of quantum light~\cite{Expanding_momentum_nature, PTC_science, PTC_amplification_science, free_electron_PTC} and emerging concepts such as photonic time crystals~\cite{lustig_PTC, Vlad_Segev_PTC, theory_application_PTC}, with direct implications for quantum computing~\cite{Moti_PTC_quantum}, all-optical computing~\cite{Alu_Non_reciprocal, Cai_alloptical}, and related next-generation technologies~\cite{theory_application_PTC}.

Recent studies have demonstrated that the optical response of ENZ systems can be tailored through hybrid material and cavity architectures. Examples include liquid-crystal-integrated ENZ nanocavities with tunable optical coupling~\cite{AOM_Liquid_Crystal}, electrically tunable dielectric singularities enabling strong Purcell enhancement in ITO hyperbolic metamaterials~\cite{SciRep_ITO_HMM}, and resonantly enhanced nonlinear and ultrafast responses in TCO-based multilayers~\cite{TCO_TPP_Negro, initial_dynamics_2}. These advances highlight the potential of structural and material design to enhance the strength, tunability, and speed of ENZ photonic systems.

Within this broader landscape, strongly confined light in an optically reconfigurable material is particularly important for achieving intense modulation on ultrafast timescales. Our previous theoretical work demonstrated that extreme electron temperatures, enabled by deeply subwavelength light confinement in transparent conducting oxides (TCOs), play a critical role in producing unconventional modulation dynamics that can approach optical-cycle timescales with strong modulation amplitude~\cite{Jaeik_TCO}. Material platforms capable of supporting such extreme confinement and electron heating are therefore of central importance to the practical advancement of emerging time-varying photonic technologies.

Particularly suitable materials for accessing such intense electron heating are the family of low-electron-density Drude (LEDD) materials. Their broadly tunable free-electron density enables a tailorable epsilon-near-zero (ENZ) regime spanning the visible to the infrared, in which light-matter interactions are greatly enhanced~\cite{Kinsey_ENZ_AIP, kinsey_NZI_Nature_review, Boyd_nonlinear_review}. These materials sustain exceptionally high electron temperatures due to their effective electron-electron interactions, enabled by reduced screening and an intrinsically low electron heat capacity, together allowing rapid thermalization~\cite{LEDD, LEDD_rigorous}. As a result, the unique physical properties associated with low electron density, combined with the ENZ effect, establish LEDD materials as promising candidates for achieving extreme light confinement and ultrafast electron-temperature dynamics. Representative LEDD materials include emerging photonic materials such as transparent conducting oxides (TCOs)~\cite{TCO_review_AOP}, transition-metal nitrides (TMNs)~\cite{Sasha_TMN, TMN_advanced_mat}, and MXenes~\cite{Mixed_MXene, DFT_MXene_ENZ}, each exhibiting ENZ behavior across distinct spectral windows: TMNs in the visible, and TCOs and MXenes across the near to far infrared. Among these classes, TCOs stand out due to their remarkably large refractive-index modulation ($\Delta n \approx 1$), recovery time several orders of magnitude faster than that of noble metals, and inherently low losses---all of which contribute to their ability to support nontrivial, femtosecond-scale modulation dynamics~\cite{Kinsey_AZO_optica, Boyd_TCO_science}.

While TCOs offer exceptional potential for ultrafast modulation, they nonetheless face the intrinsic trade-off between modulation strength and relaxation speed. Pushing toward ultrathin films amplifies electromagnetic confinement and the absorbed electron energy density, which modulate the system, but can also prolong the return to equilibrium~\cite{Hohlfeld_ballistic}. Overcoming this trade-off requires mechanisms that accelerate relaxation without sacrificing modulation depth. Interfacial carrier transfer can provide a natural route by opening additional hot electron relaxation pathways. However, the dynamical consequences of interfacial carrier exchange are not yet fully understood. Recent experimental studies have shown that carrier transport across heterointerfaces can substantially accelerate the transient response~\cite{Cai_TCO_acceleartion, MXene_acceleration}. However, a systematic theoretical understanding linking interfacial transport to optical dynamics is still lacking. Developing this connection would enable predictive interface engineering and open a broader design space for tailoring dynamic optical responses.

Building on these considerations, we investigate interfacial engineering as a central design principle for tailoring the dynamic response of TCO-based systems while preserving large modulation amplitudes. We develop a multilayer inverse design framework that concentrates light in an ultrathin TCO layer beyond what is readily achievable with intuitive cavity designs, and demonstrate that the surrounding material environment can be engineered to shape the dynamic response. A machine-learning-based two-step optimization ensures robustness to the structural variations that strongly influence the optical response in ultrathin TCOs. Direct comparisons with conventional TCO cavities highlight the advantages of inverse design in delivering both stronger and faster modulation. Finally, we show that heterointerface engineering offers a practical route to push TCO modulation into the optical-cycle regime, where previously unobserved phenomena, such as time reflection and momentum-bandgap formation in the optical domain, may become experimentally accessible.

\section{Results}\label{sec2}
\subsection{Design concept and framework}\label{sec2_1}

\begin{figure}[htbp]
    \centering
    \includegraphics[width=0.98\textwidth]{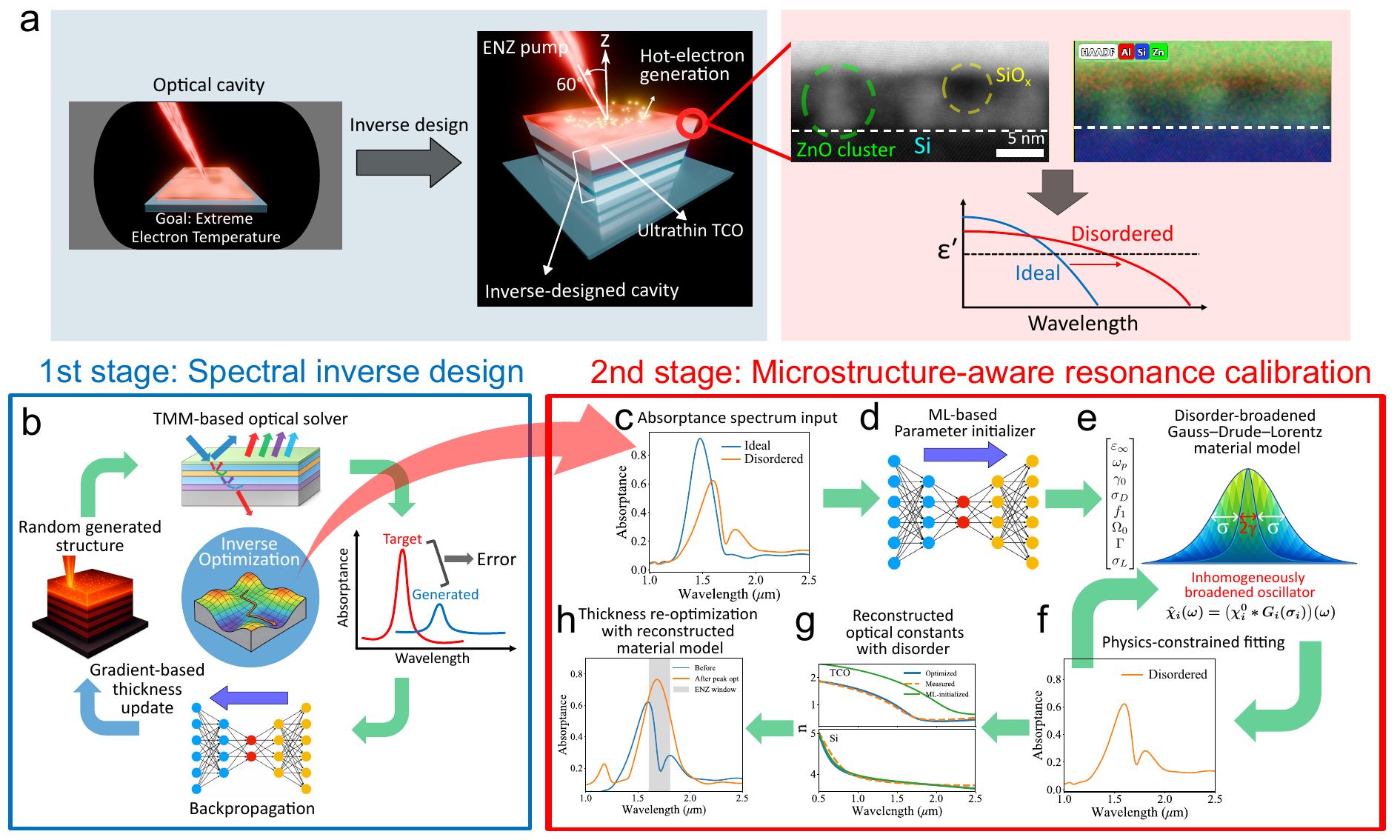}
    \caption{
    \textbf{Two-stage inverse-design framework for structural disorder-aware ENZ cavity engineering.}
    A hierarchical optimization strategy enables strong optical confinement in ultrathin TCOs while compensating for structural disorder-induced deviations in realistic materials.
    \textbf{(a)} Conceptual overview of the framework. An inverse-designed optical cavity enhances ENZ-driven hot-electron generation in an ultrathin TCO layer. At the same time, structural disorder at the TCO/Si interface perturbs the ideal permittivity and shifts the resonance.
    \textbf{(b)} First-stage spectral inverse design using a transfer-matrix-based solver with gradient backpropagation. Starting from random multilayer structures, layer thicknesses are iteratively updated through a gradient-based optimization to match a target absorptance spectrum.
    \textbf{(c)} Deviation between ideal and disordered absorptance spectra, illustrating resonance shift and degradation induced by the structural disorder.
    \textbf{(d)} Neural-network-based parameter initializer that maps the perturbed optical response to an initial estimate of effective Gauss-Drude-Lorentz (GDL) oscillator parameters.
    \textbf{(e)} Illustration of the GDL material model incorporating inhomogeneous broadening to capture structural disorder-induced variations in the permittivity of the TCO and adjacent Si layer.
    \textbf{(f)} Physics-constrained fitting procedure that refines the GDL oscillator parameters by fitting to the disorder-perturbed absorptance spectra, reconstructing the effective permittivity of the disordered TCO/Si system.
    \textbf{(g)} Reconstructed refractive index of the TCO and Si layers using the extracted GDL parameters, demonstrating accurate recovery of the disordered optical response.
    \textbf{(h)} Final re-optimization of the dielectric-layer thicknesses using the extracted optical constants under structural disorder, restoring the target resonance and absorptance under realistic material variations.
    }
  \label{fig:fig_1}
\end{figure}

Ultrathin TCO layers provide a powerful platform for achieving extreme optical confinement when their intrinsic limitations are properly addressed. In the ENZ regime, strong field enhancement and slow-light effects substantially amplify light–matter interactions. When combined with the optical confinement enabled by the ultrathin layer thickness, these effects lead to strong absorption and intense electron heating. Pushing TCOs into the ultrathin regime introduces two central challenges. First, the high intrinsic transparency of TCOs reduces absorptance as film thickness decreases, thereby limiting optical confinement. This constraint can be mitigated by embedding the ultrathin TCO layer within a cavity geometry (Figure~\ref{fig:fig_1}(a), left)~\cite{TCO_TPP, TCO_TPP_Negro, TCO_TPP_Negro_2}. Second, ultrathin TCOs are susceptible to structural disorder induced by heavy doping and interfacial instability, which perturb the permittivity from its ideal response. These deviations include interfacial oxide formation, localized amorphization, dopant segregation, and stoichiometric variations at the TCO/Si boundary. A representative aluminum-doped zinc oxide (AZO)/Si interface was examined using scanning transmission electron microscopy (STEM) and energy-dispersive spectroscopy (EDS), revealing pronounced interfacial disorder (Figure~\ref{fig:fig_1}(a), right). The observed disorder consists of compositional separation into $\mathrm{SiO_2}$, ZnO clusters, and Al-enriched regions, together with partial amorphization. The Al-enriched region is visible as the red layer in the EDS map overlaid on the STEM image. While such distortions are negligible in thicker layers of hundreds of nanometers, they become dominant when the film thickness approaches the scale of the disordered interfacial region, which is precisely the ultrathin regime explored in this study.

Simultaneously overcoming transparency-limited confinement and structural disorder-induced shifts in optical properties requires a design strategy that is both robust and adaptable. To enable these capabilities, we introduce a multilayer inverse-design framework that enhances absorption in ultrathin TCOs while preserving robustness against interfacial imperfections through an integrated disorder-compensation loop. Figure~\ref{fig:fig_1}(b--h) illustrate the overall flow of this multi-stage inverse-optimization framework. In this framework, the static and dynamic optical responses are analyzed using the first-stage optimized cavity structures, which represent the ideal theoretical designs. A subsequent fabrication-aware optimization step bridges the ideal designs and fabricable geometries by explicitly accounting for structural disorder. The first stage maximizes optical confinement and absorption, while the second preserves this optimized resonance response under realistic material disorders, facilitating experimental realization of the theoretical design.

The first-stage inverse design approach establishes a cavity-assisted ENZ resonance that overcomes the intrinsically low absorptance of ultrathin TCO films. The optimization is performed using a transfer-matrix optical solver with gradient backpropagation, yielding the optimal layer thicknesses and TCO carrier concentration for a target absorptance profile. The dielectric function of the TCO is modeled using a carrier-density-dependent Drude formulation (Eq.~\ref{eq:Drude}). Although demonstrated using TCOs, this design strategy is broadly applicable to other Drude-type materials, including transition-metal nitrides, MXenes, and conventional plasmonic metals. The framework can also accommodate physics-driven constraints, such as fixing the TCO thickness to control the absorbed energy density or enforcing material combinations that promote faster relaxation.

Figure~\ref{fig:fig_1}(b) schematically summarizes the first-step inverse-design workflow, which is based on a differentiable transfer-matrix-method framework similar to those used in previous multilayer optical-design studies~\cite{Caldwell_design_NM, Omer_design}. The multilayer cavity is composed of an ultrathin TCO layer and a five-layer alternating Si/$\mathrm{SiO_2}$ dielectric stack designed to support a cavity resonance. In the inverse-design loop, the Si/$\mathrm{SiO_2}$ layer thicknesses and TCO carrier concentration are randomly initialized and passed to a differentiable transfer-matrix method (TMM) solver to calculate the spectral absorptance of the multilayer stack. After this step,  the generated spectrum is compared with the target response using a mean-squared-error loss, $\mathcal{L}=\frac{1}{N_{\lambda}}\sum_{\lambda}\left|A_{\mathrm{target}}(\lambda)-A_{\mathrm{calc}}(\lambda)\right|^2$, and the resulting gradients are backpropagated through the TMM solver to update the design parameters, namely the dielectric layer thicknesses and TCO carrier concentrations. Iterative refinement converges to an optimized multilayer cavity that realizes the desired spectral response.

The second stage compensates for the inevitable structural disorder in TCO-based multilayer cavities that shifts the resonance from its target value. This step is critical for translating the optimized design into a practical realization, ensuring that the fabricated device preserves the intended resonance characteristics. As illustrated in Figure~\ref{fig:fig_1}(c--h), the realistic absorptance spectrum of the fabricated stack is incorporated into a regression-based correction loop to infer the effective permittivity of the TCO and the adjacent Si layer, accounting for realistic interfacial disorder. The inferred effective permittivity then guides the re-optimization of the dielectric layer thicknesses to restore the target resonance. This correction step is especially useful for devices subjected to repeated deposition and thermal processing, where continuous structural evolution makes independent layer-by-layer ellipsometric characterization and thickness verification difficult.

The red box encompassing Figure~\ref{fig:fig_1}(c--h) summarizes the second-stage optimization loop, from disorder-induced resonance degradation to its recovery. Figure~\ref{fig:fig_1}(c) first compares the response of the ideal and disordered cavities. The ideal absorptance is calculated from the permittivity of a 238-nm-thick gallium-doped zinc oxide (GZO) film, assuming that its optical constants remain unchanged as the thickness is reduced to the ultrathin regime. The disordered case instead uses the permittivity measured from the actual ultrathin film deposited under the same conditions, which exhibits pronounced interfacial disorder. The resulting spectral shift and the decrease in peak absorptance highlight the impact of structural variations and motivate the second-stage correction.

The non-ideal (disordered) absorptance spectrum is used as input to a neural-network-based parameter extractor (Figure~\ref{fig:fig_1}(d)), which outputs sixteen parameters as an informed initial guess for Gauss-Drude-Lorentz (GDL) fitting (Figure~\ref{fig:fig_1}(e)). This ML-generated parameter set partially captures the measured complex refractive index $(n, k)$ and serves as an effective starting point for the subsequent gradient-based extraction of the effective $n$, $k$ values of the disordered material (Figure~\ref{fig:fig_1}(f)). Despite the high-dimensional 16-parameter fitting space, the optimized refractive index shows close agreement with the ellipsometry measurements (Figure~\ref{fig:fig_1}(g)), demonstrating the robustness of the framework. Finally, re-optimizing the dielectric stack with the extracted optical constants restores the resonance condition and yields an approximately $\sim$50\% enhancement in absorptance (Figure~\ref{fig:fig_1}(h)). Overall, this second-stage loop accounts for realistic structural disorder and restores the desired optical resonance peak of the inverse-designed cavity.

The core of the second optimization step is to extract the effective optical properties of the disordered TCO/Si bilayer (Figure~\ref{fig:fig_1}(g)). This is achieved by parameterizing the disorder-modified refractive indices of the TCO and adjacent Si layers using Gauss-Drude and Gauss-Lorentz oscillators, which introduce Gaussian broadening into the conventional Drude-Lorentz formalism~\cite{Lucy_broadening}. This inhomogeneous broadening provides a compact description of structural disorder-induced variations~\cite{Lucy_broadening}, including dopant segregation, stoichiometric fluctuations, and interfacial oxide formation. The broadened oscillator parameters are then fitted to the measured absorptance spectrum to reconstruct the effective refractive index profiles of the disordered TCO and Si layers. In this way, the model provides a physically interpretable representation of the disorder-modified TCO/Si bilayer embedded in the TCO-integrated multilayer cavity.

The corresponding total permittivity of the TCO and Si layer, parameterized by the GDL oscillators, is written as
\begin{equation}
\varepsilon(\omega)=\varepsilon_\infty+\chi_{\mathrm{GD}}(\omega)+\chi_{\mathrm{GL}}(\omega),
\label{eq:GDL}
\end{equation}
where $\varepsilon_\infty$ is the background permittivity and $\omega=2\pi c/\lambda$ is the angular frequency.
The conventional Lorentz (Drude) models implicitly assume homogeneous (lifetime-limited) broadening, for which the ideal oscillator (conductivity) is broadened by a Cauchy probability distribution, giving rise to the familiar damping parameters $\Gamma$ and $\gamma$,
\begin{equation}
\chi_{\mathrm{D}}(\omega;\Gamma)= -\frac{\omega_\text{p}^2}{\omega^2+\ii\Gamma\omega}, \qquad
\chi_{\mathrm{L}}(\omega;\Omega)=\frac{f_1}{\Omega^2-(\omega+\ii\gamma)^2},
\label{eq:DL}
\end{equation}
where $\omega_\text{p}$ is the plasma frequency, $\Gamma$ is the free-carrier scattering rate, $f_1$ is the oscillator strength, $\Omega$ is the resonance frequency, and $\gamma$ is the Lorentz damping rate.

Following our probability-based formulation of dispersion developed in Ref.~\cite{Lucy_broadening}, we generalize this homogeneous broadening by replacing the underlying Cauchy profile with a Voigt profile, obtained by convolving it with a Gaussian distribution that accounts for disorder-induced inhomogeneous broadening. The resulting generalized Gauss-Drude and Gauss-Lorentz susceptibilities are conveniently expressed in terms of the Faddeeva function (details are provided in the Supplementary Information),
\begin{equation}
\begin{aligned}
\chi_{\rm GD}(\omega)
&=
\frac{\ii\omega_{\rm p}^{2}\sqrt{\pi}}
     {\omega\sigma_{\rm D}\sqrt{2}}\,
w\!\left(
\frac{\omega+\ii\Gamma}
     {\sigma_{\rm D}\sqrt{2}}
\right),\\[2mm]
\chi_{\rm GL}(\omega)
&=
\frac{\ii f_1\sqrt{\pi}}
     {2\Omega\sigma_{\rm L}\sqrt{2}}
\left[
w\!\left(
\frac{\omega+\ii\gamma-\Omega}
     {\sigma_{\rm L}\sqrt{2}}
\right)
-
w\!\left(
\frac{\omega+\ii\gamma+\Omega}
     {\sigma_{\rm L}\sqrt{2}}
\right)
\right],
\end{aligned}
\label{eq:Convolution}
\end{equation}
where $\sigma_\mathrm{D}$ and $\sigma_\mathrm{L}$ quantify the degree of inhomogeneous broadening. The permittivity of each layer is therefore described by eight unknown GDL parameters,
\begin{equation}
\{\varepsilon_\infty,\omega_p,\Gamma,\sigma_\mathrm{D},f_1,\Omega,\gamma,\sigma_\mathrm{L}\},
\label{eq:param_array}
\end{equation}
resulting in a total of sixteen parameters for the TCO and Si layers. These parameters, together with their material-specific bounds, are mapped to the corresponding optical absorptance spectra using the TMM solver.

The neural-network estimator is then trained on 50{,}000 simulated samples to map the cavity optical absorptance spectra onto the GDL parameter set defined in Eq.~\ref{eq:param_array}, thereby providing informed initial guesses for the subsequent optimization. This ML-generated initialization helps navigate the high-dimensional parameter space of the broadened oscillator model. It provides a physically meaningful starting point for extracting the disorder-modified refractive index through nonlinear least-squares fitting. The extracted refractive indices of the disordered TCO and Si layers are then used in a final multilayer re-optimization step to adjust the dielectric layer thicknesses and restore high absorptance under realistic structural disorder.

For experimental implementation, the inverse problem can be more constrained by jointly fitting reflectance and transmittance spectra measured at multiple angles and polarizations. In particular, variable-angle spectroscopic ellipsometry (VASE) measurements of the parameters $\Psi$ and $\Delta$ provide additional polarization-dependent amplitude and phase information that can be incorporated directly into the transfer-matrix-based optimization. Each additional measurement provides an independent constraint, reducing the likelihood that different GDL parameter sets reproduce the same measured responses simultaneously. Measurement noise may also challenge thickness re-optimization. However, its influence can be largely mitigated by performing multiple measurements and averaging the measured responses. Incorporating the VASE measurements of $\Psi$ and $\Delta$ together with noise-aware calibration therefore represents an important extension of the present framework and will be pursued in future experimental studies.

\subsection{Static Optical Response of Inverse-Designed TCO Cavities}

\begin{figure}[htbp]
    \centering
    \includegraphics[width=1\textwidth]{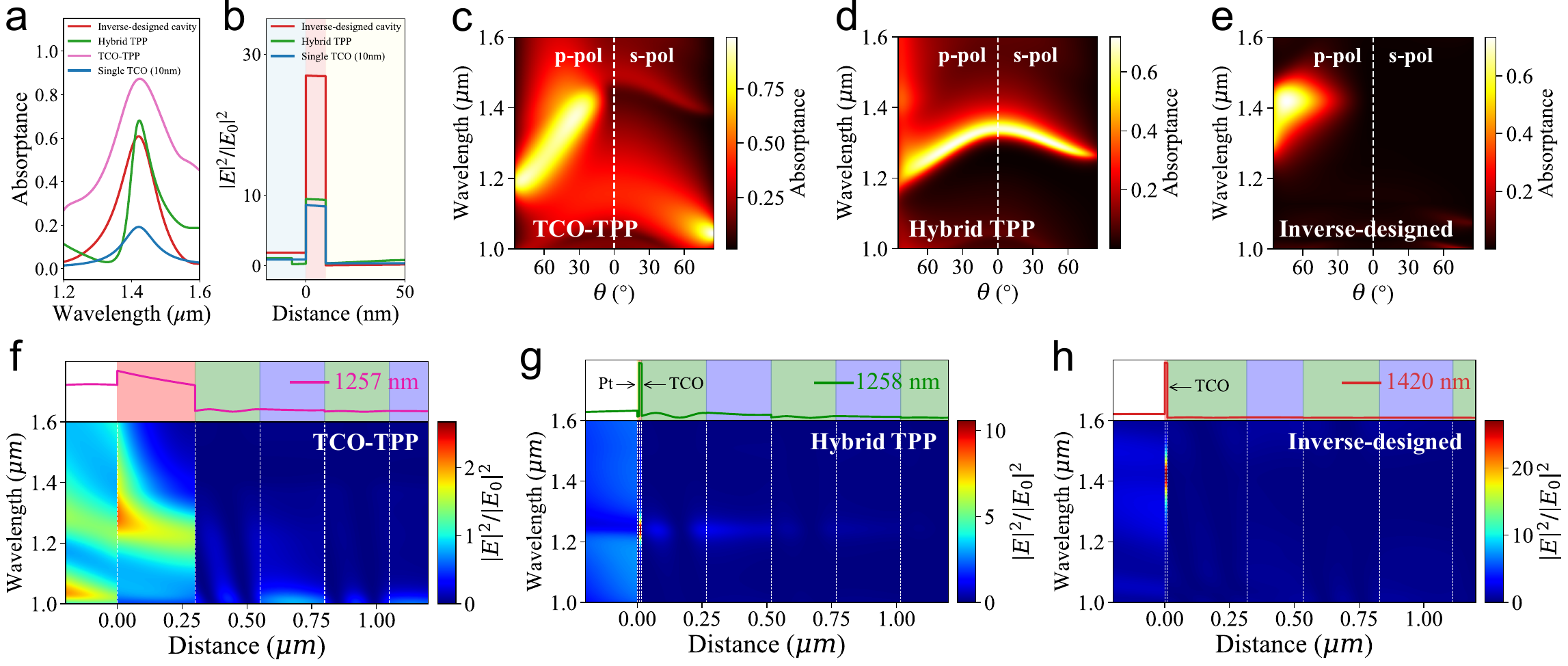}
    \caption{
    \textbf{Comparison of optical absorption and field confinement across conventional and inverse-designed TCO cavities.}
    The inverse-designed cavity enhances ENZ-driven absorption and enables superior field confinement within an ultrathin TCO layer compared to conventional TPP-based designs.
    \textbf{(a)} Calculated absorptance spectra of a single TCO film, TCO-TPP, hybrid TPP, and inverse-designed cavities, highlighting substantially enhanced absorption in the cavity geometries relative to the single film.
    \textbf{(b)} Field localization within the TCO layer at the ENZ wavelength for the different geometries, with the strongest confinement observed in the inverse-designed cavity.
    \textbf{(c)--(e)} Angle-resolved absorptance spectra for the TCO-TPP, hybrid TPP, and inverse-designed cavities, respectively. The TPP-based structures exhibit a characteristic parabolic dispersion, whereas the inverse-designed cavity maintains broadband, angle-robust absorption near the ENZ region.
    \textbf{(f)--(h)} Corresponding spatial distributions of normalized electric field intensity $|E|^2/|E_0|^2$ at a representative incidence angle ($60^\circ$) for the TCO-TPP, hybrid TPP, and inverse-designed cavities, respectively. The inverse-designed structure exhibits the strongest confinement within the ultrathin TCO layer. At the same time, the TCO-TPP and hybrid TPP configurations show reduced localization due to spatially distributed absorption, and loss in the metal layer, respectively.
    }
  \label{fig:fig_2}
\end{figure}

We evaluate the performance advantages enabled by the inverse design framework, beginning with its static optical response. The inverse-designed cavity yields over a threefold increase in absorptance near the ENZ region relative to a single TCO film of identical thickness on fused silica (Figure~\ref{fig:fig_2}(a)), while also producing a threefold enhancement of the local electric-field intensity (Figure~\ref{fig:fig_2}(b)). To further examine the enhancement, we compare the inverse-designed structure with conventional, intuitively designed cavities---specifically, two representative TCO-Tamm plasmon polariton (TPP) configurations, with and without an added plasmonic metal layer.

TPP is a standing surface mode localized at the interface between a plasmonic layer and a reflective photonic crystal~\cite{TPP_2006, TPP_2007, TPP_2008}. Its formation requires a balance between absorption within the plasmonic medium and radiative leakage, such that the plasmonic layer effectively functions as a semi-transparent mirror~\cite{TCO_TPP}. In metal-free configurations, the TCO film must itself provide the plasmonic response, which typically necessitates a relatively thick ($\sim$300\,nm) TCO layer, whose ENZ wavelength coincides with the intended TPP wavelength. In contrast, incorporating a plasmonic metal layer relaxes this requirement: the metal’s stronger metallicity satisfies the coupling condition at much smaller thickness, allowing the TCO film to remain ultrathin as a spacer layer between the metal and photonic crystal. In such hybrid designs, the metal establishes the TPP mode, and the resulting field couples into the adjacent TCO spacer layer, producing strong hybrid confinement when the TPP resonance overlaps with the ENZ band. In this work, a 7\,nm Pt layer is employed to achieve robust localization of the TPP mode within the TCO ENZ regime.

Absorption within the TPP structure is substantially enhanced through cavity-assisted resonances (Figure~\ref{fig:fig_2}(a)), reflected by the TCO–TPP structure achieving nearly 90$\%$ absorptance. Notably, both the hybrid TPP and inverse-designed cavities exceed 60$\%$ absorptance while employing only a 10\,nm TCO layer, representing more than a threefold enhancement compared to a single film of identical thickness. The angle-resolved absorptance spectra in Figure~\ref{fig:fig_2}(c--e) further highlight the distinct resonance behavior exhibited by each cavity geometry. The TPP structures exhibit a characteristic parabolic dispersion of the TPP mode near the ENZ wavelength, indicating a resonance that is strongly dependent on incident angle. By contrast, the inverse-designed cavity sustains a broad, high-absorptance response near the ENZ band, even at large incidence angles, revealing a resonance governed primarily by the ENZ regime rather than by plasmon-induced effects. The E-field distributions at the peak absorption wavelength under 60$^\circ$ incidence are shown in Figure~\ref{fig:fig_2}(f--h) overlaid on the corresponding structures, further highlighting the distinct confinement behavior of each structure.

Despite the strong absorptance enhancement achieved across all cavity geometries, the inverse-designed cavity exhibits distinctly superior light confinement within the TCO layer, as evident from the comparison in Figure~\ref{fig:fig_2}(b). The inverse-designed cavity delivers roughly threefold higher local electric-field intensity than both the hybrid TPP and the single-layer film, and nearly a 15-fold increase in field intensity compared to the TCO-TPP structure. The reduced confinement in the TCO-TPP cavity originates from the relatively thick (300\,nm) TCO layer. Despite near-unity absorptance, the absorbed energy is spread across a substantially larger volume (Figure~\ref{fig:fig_2}(f)), diminishing local field intensity. The hybrid TPP design improves localization compared to the metal-free TCO-TPP configuration due to the ultrathin TCO thickness. However, the TPP mode couples not only to the ENZ resonance of the TCO but also to the lossy Pt layer, leading to substantial absorption within Pt and limiting the field intensity achievable in the TCO (Figure~\ref{fig:fig_2}(g)). As a result, the field confinement within the TCO region remains comparable to that of a single ultrathin film, even though the total absorption is strongly enhanced. By contrast, the inverse-designed structure concentrates the field directly within an ultrathin TCO layer without relying on a plasmonic metal, enabling substantially more efficient light localization through cavity-enhanced ENZ effects (Figure~\ref{fig:fig_2}(h)). Notably, the field enhancement in the optimized cavity is spectrally aligned with the ENZ condition, unlike in the TPP cavity, where strong dispersion shifts the resonance away from the ENZ wavelength at higher incidence angles.

\subsection{Dynamic Optical Response of Inverse-Designed TCO Cavities}

\begin{figure}[htbp]
    \centering
    \includegraphics[width=1\textwidth]{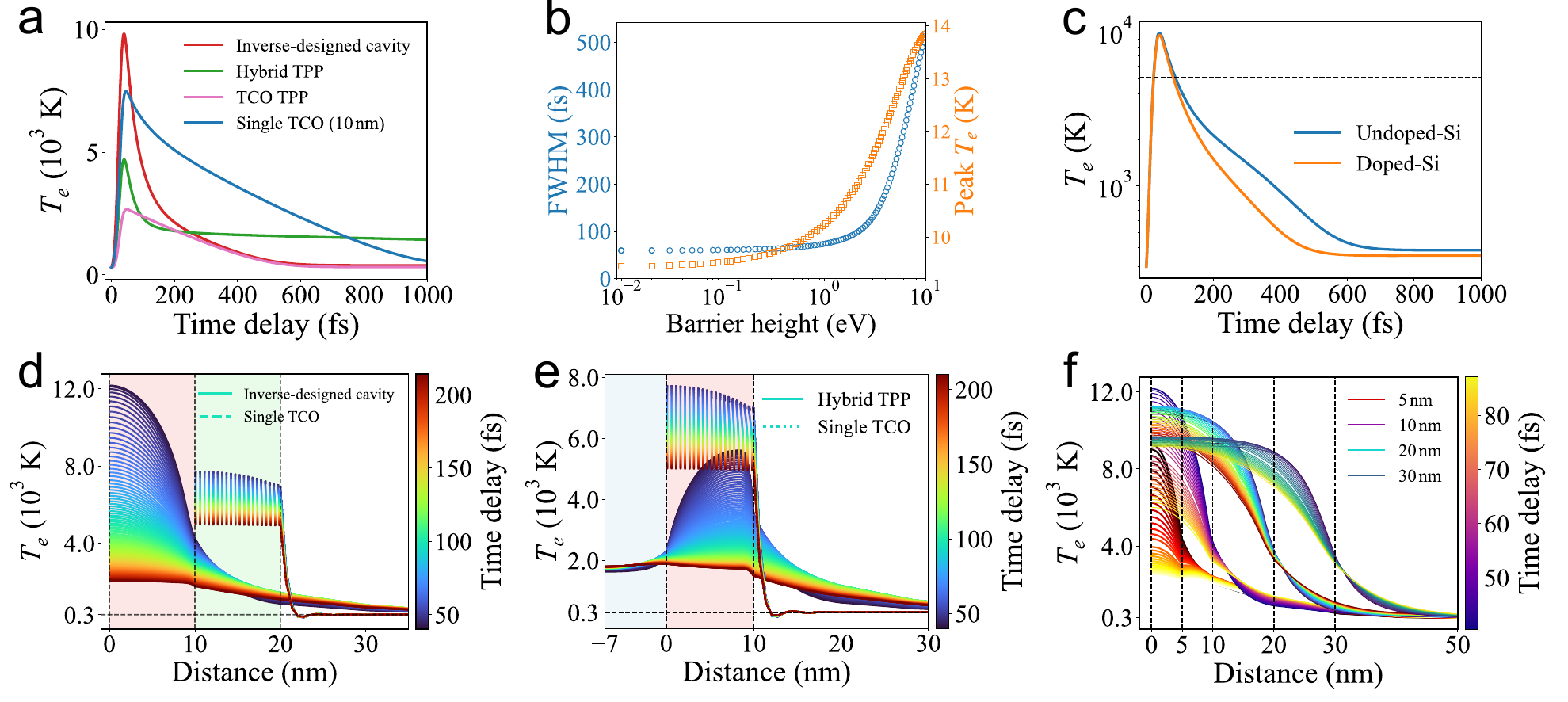}
    \caption{
    \textbf{Electron temperature dynamics and interfacial engineering in an inverse-designed TCO cavity.}
    The inverse-designed cavity simultaneously achieves higher peak electron temperature and faster relaxation through combined ENZ-enhanced confinement and efficient interfacial carrier extraction. In addition, TCO thickness and interfacial conditions provide additional control over the dynamic response.
    \textbf{(a)} Temporal evolution of electron temperature in the TCO for a single TCO film, TCO-TPP, hybrid TPP, and inverse-designed cavity geometries, showing enhanced peak temperature and faster decay in the inverse-designed structure.
    \textbf{(b)} Dependence of relaxation time (FWHM) and peak electron temperature on the TCO/Si interfacial barrier height, illustrating the trade-off between modulation strength and relaxation speed.
    \textbf{(c)} Electron temperature dynamics in the TCO for undoped and heavily doped adjacent Si layer, highlighting suppression of the long-lived relaxation tail with increased doping.
    \textbf{(d),(e)} Spatiotemporal evolution of electron temperature for the inverse-designed and hybrid TPP cavities, respectively. The inverse-designed structure exhibits faster relaxation and stronger spatial temperature gradients due to efficient carrier extraction, whereas the hybrid TPP structure shows a rapid drop in peak electron temperature induced by multiple spatial relaxation channels.
    \textbf{(f)} Spatiotemporal electron temperature profiles in inverse-designed cavities individually optimized for each TCO thickness, demonstrating thickness-dependent control of relaxation dynamics.
    }
    \label{fig:fig_3}
\end{figure}

The electron temperature dynamics of the inverse-designed cavity can be markedly enhanced through interfacial engineering. Using a two-temperature model with modified interfacial boundary conditions, we calculate the spatiotemporal evolution of the electron temperature to elucidate the role of the interface in the dynamic response. This analysis clarifies the origin of the ultrafast, strong modulation in the inverse-designed cavity, explains the weaker response and long-lived tail of the hybrid TPP cavity, analyzes the effects of barrier height and Si doping on carrier relaxation, and examines the influence of TCO thickness on the overall dynamics.

\subsubsection*{Enhanced Modulation Dynamics in TCOs Enabled by Inverse Design}
Figure~\ref{fig:fig_3}(a) shows that the inverse-designed cavity attains a fast and strong electron temperature modulation. The peak electron temperature is enhanced by $\sim$4$\times$ relative to the TCO-TPP structure and $\sim$2$\times$ relative to the hybrid TPP design. In addition, the inverse-designed cavity exhibits the fastest relaxation dynamics, as quantified by the full width at half maximum (FWHM) of the electron temperature curve. Notably, the inverse-designed cavity exhibits more than a 5-fold reduction in FWHM compared to the single-layer film and approximately a 4-fold reduction relative to the TCO–TPP structure. Although its FWHM is comparable to that of the hybrid TPP cavity, it achieves nearly twice the peak electron temperature. The FWHM and peak electron temperature of the four geometries are summarized in Table~\ref{tab:performance}. This improvement in the inverse-designed cavity arises from the simultaneous realization of tight light confinement near the ENZ regime and efficient carrier extraction into the adjacent Si layer. The intuitively designed cavities, by comparison, exhibit relatively modest, slower modulation despite their high absorptance. In particular, the hybrid TPP structure exhibits a pronounced relaxation tail after 150\,fs which limits its relaxation rate.

\begin{table}[H]
\centering
\caption{Peak electron temperature and relaxation time (FWHM) for the four TCO-based geometries.}
\label{tab:performance}
\begin{tabular}{lcc}
\hline
Geometry & Peak $T_e$ ($\times 10^3$ K) & FWHM (fs) \\
\hline
Inverse-designed & 10 & 68 \\
Single TCO & 8 & 352 \\
TCO-TPP & 2.5 & 269 \\
Hybrid TPP & 5 & 71 \\
\hline
\end{tabular}
\end{table}

The comparatively weak and slow modulation observed in the intuitively designed cavities underscores the critical influence of the surrounding material environment in the TCO dynamic response. A particularly counterintuitive observation is that the hybrid TPP cavity, despite achieving more than a 3$\times$ enhancement in absorptance within a 10\,nm TCO film, produces a lower peak electron temperature than the single TCO film. As discussed in more detail in the subsequent spatio-temporal electron temperature analysis, this behavior arises from extremely rapid electron temperature relaxation in TCO through the Pt layer, which prevents substantial temperature buildup. In comparison, the single TCO film on fused silica confines hot electrons within the TCO layer. The large interfacial energy barrier at the TCO/fused silica interface suppresses carrier transport, leading to a higher peak electron temperature. However, because the electrons remain localized within an ultrathin region and relax exclusively through lattice cooling (electron-phonon coupling), without any spatial extraction pathway, the subsequent relaxation becomes significantly slower.

\subsubsection*{Interfacial Engineering of Ultrafast Relaxation}
The interfacial barrier height strongly influences the dynamic response by governing the efficiency of hot-carrier relaxation via the spatial transport pathways. This behavior is captured by the two-temperature model through a thermionic boundary condition at the TCO/Si interface (Eq.~\ref{eq:Interfacial_BC})~\cite{Thermionic_Sim, Thermionic_space_charge}. Both the relaxation time (FWHM) and peak electron temperature exhibit a strong dependence on barrier height, both increasing as the interface becomes more insulating (Figure~\ref{fig:fig_3}(b)). Notably, the relaxation time remains nearly constant up to a barrier height of $\sim$1\,eV before increasing sharply, whereas the peak temperature rises more gradually. This disparity enables enhancement of modulation strength without a proportionate penalty in response speed. Because the barrier height can be readily tuned through deposition conditions, interface treatments, annealing, and doping, it provides a versatile, yet underexplored parameter for engineering ultrafast carrier dynamics.

The TCO/Si interfacial barrier is experimentally tailorable through deposition and interface engineering. For example, oxygen content during TCO deposition modifies its work function and Schottky barrier height, while surface treatment and annealing further control these parameters through the formation and passivation of interfacial SiO$_x$ and interface states~\cite{Zhang_SciAdv_2023,Li_Materials_2018,Gao_ACSAMI_2017,Zhang_ADFM_2022}. If a wider tuning range or improved stability is required, a deliberately deposited ultrathin dielectric can be used to control the effective Schottky barrier height~\cite{Yao_JACS_2016}. The experimentally realized barrier can be extracted using current--voltage and capacitance--voltage measurements. The range accessible to a TCO/undoped-Si junction is typically on the order of 0.45--1\,eV. Accordingly, the larger effective barriers in Figure~\ref{fig:fig_3}(b) can be achieved through a strongly blocking interface by introducing a dielectric interlayer.

Moreover, reducing the interfacial barrier width via heavy Si doping can enhance the relaxation dynamics via Ohmic junction formation~\cite{TCO_Si_Ohmic}, effectively suppressing the barrier-limited relaxation tail at low electron temperatures (Figure~\ref{fig:fig_3}(c)). When doping reduces the depletion width to only a few nanometers through strong internal band bending, carrier transport transitions from thermionic emission to field emission (tunneling), effectively removing the barrier limitation and driving the interface toward Ohmic behavior. According to thermionic–field-emission theory, this transition occurs at a Si carrier concentration of approximately $6 \times 10^{18}\,\mathrm{cm}^{-3}$ in our geometry~\cite{Thermionic_field_theory}. The corresponding electron-temperature dynamics of the inverse-designed cavity with and without heavy doping in the interfacing Si layer (plotted on a logarithmic scale in Figure~\ref{fig:fig_3}(c)) reveal a clear divergence in the low-temperature regime. Although the peak modulation remains similar in both cases, the traces begin to separate after $\sim$150\,fs, when the temperature decreases to approximately half of its maximum value, as indicated by the dotted lines. These results demonstrate that heavy doping substantially suppresses the long-lived background tail. This feature is particularly advantageous for applications requiring controlled and repeatable temporal modulation, such as photonic time crystals, optical neural networks, and ultrafast optical switches/modulators~\cite{Nature_Physics_PTC, Alu_nonlocal_PTC, Cai_alloptical}.

\subsubsection*{Mechanisms Governing Fast and Slow Relaxation in Cavity-Integrated TCOs}
A spatially resolved electron temperature distribution and its temporal evolution are obtained from the two-temperature model, incorporating thermionic boundary conditions at the interface. This spatiotemporal electron-temperature analysis of the inverse-designed cavity reveals that its rapid optical response arises from efficient hot-carrier transfer into the adjacent material. As shown in Figure~\ref{fig:fig_3}(d), the electron temperature profile within the cavity (solid lines) develops a pronounced spatial gradient across the TCO layer, in stark contrast to the nearly uniform distribution observed in the single-film reference (dotted lines). This gradient indicates that thermionic emission at the interface not only alters the local boundary temperature but also modifies energy distribution throughout the entire film once thermalization is established---typically within 10--20\,fs in TCOs due to rapid electron-electron scattering~\cite{TTM_FDTD, LEDD_rigorous}. A detailed discussion of the thermalization time is provided in our previous work~\cite{Jaeik_TCO}. Consequently, the TCO relaxation dynamics become more sensitive to the surrounding material environment and cavity geometry than previously anticipated.

Although spatial relaxation accelerates TCO dynamics, excessive relaxation pathways can produce undesirable long-lived tails once the temperature profile becomes spatially uniform. Here, we revisit the hybrid TPP cavity response to reveal the physical origin of the long-lived relaxation tail in Figure~\ref{fig:fig_3}(a) (after $\sim$150\,fs in the green curve). This type of response has been observed in various experimental studies, but a clear physical explanation has been lacking~\cite{Cai_TCO_acceleartion, Soham_Fast_Slow, MXene_acceleration, initial_dynamics_1, initial_dynamics_2}. In the hybrid TPP cavity, the Pt/TCO interface forms an Ohmic contact, while the TCO/Si interface presents a relatively low Schottky barrier. Carrier transport is therefore permitted at both boundaries, allowing hot electrons to drain rapidly into the surrounding layers. This dual extraction suppresses temperature buildup in the TCO and quickly flattens the spatial temperature profile after $\sim$150\,fs (Figure~\ref{fig:fig_3}(e)). Once this uniform state is established, carriers begin to flow back from the Pt layer into the TCO because the TCO cools much more rapidly than Pt (with an order-of-magnitude smaller $\tau_{ep} \sim C_e / G_{ep}$), creating a reversed temperature gradient. The resulting backflow partially replenishes the hot electrons and slows the net cooling in the TCO layer, giving rise to the long-lived relaxation tail observed after $\sim$150\,fs. This mechanism demonstrates that the inverse-designed cavity benefits from strong optical confinement with only a single active interface, enabling controlled spatial relaxation without introducing slow background tails.

In addition to interfacial effects, the TCO thickness also governs carrier dynamics, a factor that has been largely overlooked. Thinner layers are more strongly influenced by interfacial transport, whereas thicker layers absorb greater optical energy. At the same time, reduced thickness enhances field confinement, establishing a trade-off between absorption volume and spatial energy localization that governs modulation strength. To systematically evaluate this interplay, we inverse-designed a series of cavities optimized to maximize ENZ absorption across TCO thicknesses ranging from 5 to 30\,nm. Figure~\ref{fig:fig_3}(f) shows that a 10\,nm layer yields the most favorable combination of peak electron temperature and relaxation speed, supported by a pronounced spatial electron temperature gradient. At 5\,nm, carriers relax too rapidly to sustain strong heating, while thicknesses exceeding 20\,nm develop a flat temperature region that delays cooling.

\begin{figure}[htbp]
    \centering
    \includegraphics[width=1\textwidth]{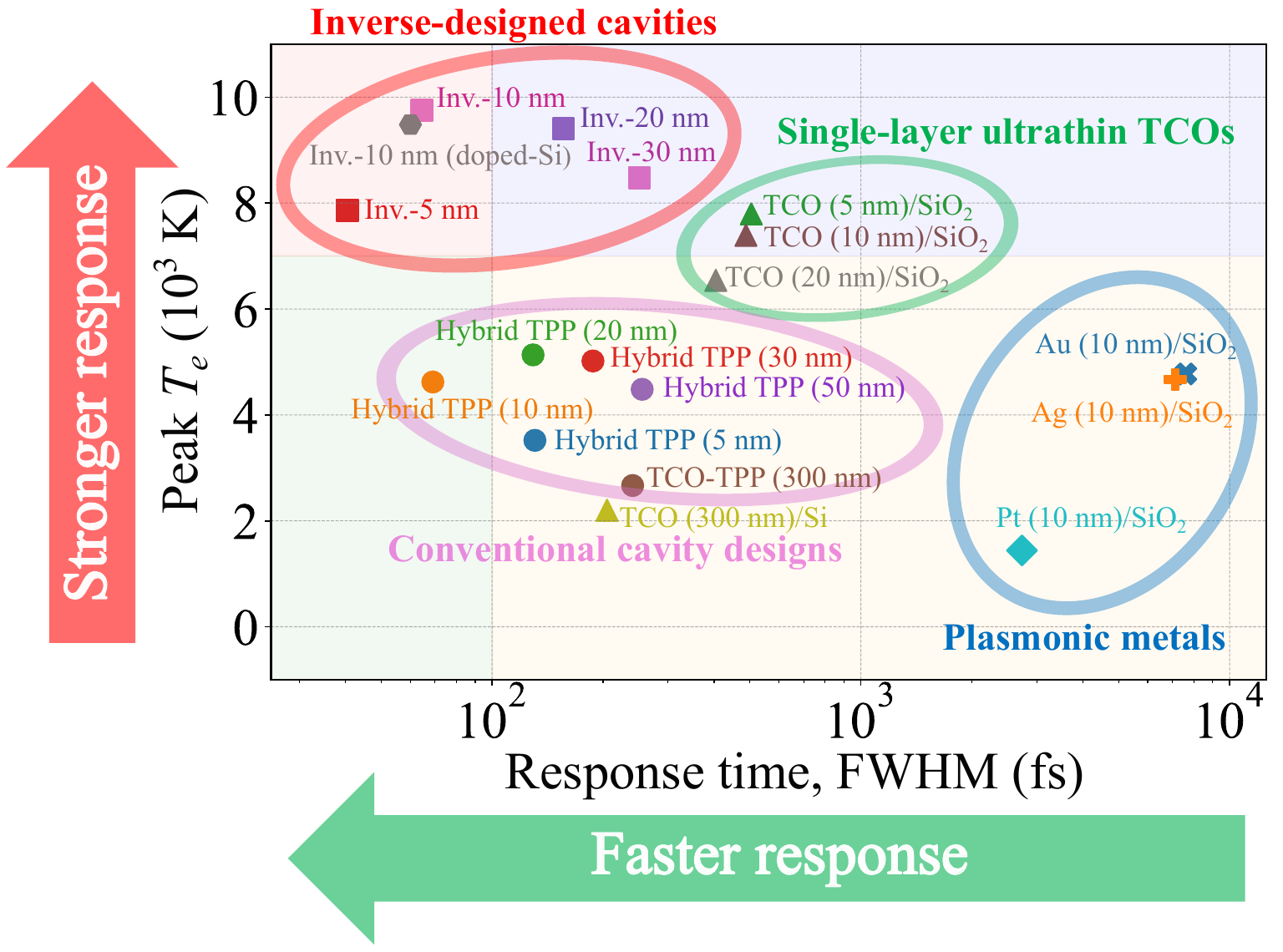}
    \caption{
    \textbf{Performance comparison of ultrafast optical modulation across different ENZ platforms.}
    Simulated peak electron temperature and relaxation time (FWHM) for various cavity geometries and materials under their corresponding ENZ pump excitation conditions.
    Inverse-designed cavities occupy the upper-left region, achieving both stronger modulation and orders-of-magnitude faster response than conventional TCO cavities and plasmonic metals.
    }
    \label{fig:fig_4}
\end{figure}

Overall, this study establishes inverse-designed TCO cavities as a distinctly suitable platform for the emerging time-varying photonic applications, outperforming a wide range of conventional TCO-based structures. Figure~\ref{fig:fig_4} summarizes the simulated dynamic responses under ENZ pumping for various material and geometric configurations. The inverse-designed cavity achieves approximately fivefold faster relaxation than single ultrathin TCO films and approximately twice the modulation strength of intuitively designed TPP cavities, while also markedly surpassing conventional plasmonic metals. These results underscore how strong ENZ confinement, combined with engineered interfacial relaxation, enables ultrafast and large-amplitude modulation unattainable with conventional materials and designs.

\subsection{Toward optical-cycle-timescale modulation via interfacial carrier relaxation}

Here, we investigate how the ultrafast electron temperature dynamics enabled by strong ENZ confinement and interface-controlled dynamics bring the system toward the optical-cycle-timescale photonic regime. While this regime has opened profound opportunities and stimulated extensive theoretical interest, the development of practical material platforms capable of supporting such rapid dynamics remains a major challenge~\cite{Expanding_momentum_nature, Moti_disordered_PTC, Vlad_Segev_PTC, Nature_Physics_PTC}. To evaluate how these dynamics manifest in experimentally measurable quantities, we compute the transient pump-probe transmittance response using Eq.~\ref{eq:transmission}, which calculates the transient dynamics from the time-dependent complex permittivity~\cite{Ncomm_del_T_Eq, TPP_modeling}. The permittivity is obtained by updating the plasma frequency and Drude damping according to the electron temperature evolution predicted by the two-temperature model (Eqs.~\ref{eq:TTM_e} and \ref{eq:TTM_L}).

Previous experimental observations of oscillatory dynamics in TCOs have attracted considerable interest~\cite{Moti_TCO, Moti_CLEO}. Our previous study provided a theoretical interpretation of the physical mechanisms that may underlie this behavior~\cite{Jaeik_TCO}. The oscillatory transient response originates from the distinctive hot-carrier physics of TCOs. Under ENZ excitation, electron heating significantly broadens the Fermi–Dirac distribution and can approach the Fermi temperature, populating states well above the Fermi level. Due to the distinct non-parabolicity of the conduction band, the electron energy redistribution dynamically modifies the free electron effective mass and shifts the plasma frequency $\omega_\mathrm{p}$, while increased phonon, impurity, and grain-boundary scattering increases the Drude damping $\Gamma$~\cite{Sivan_damping, TCO_scattering_Nanophotonics}. The concurrent evolution of the real and imaginary components of the permittivity, driven by simultaneous changes in $\omega_\mathrm{p}$ and $\Gamma$, produces an oscillatory transient optical response~\cite{Jaeik_TCO}. In ultrathin TCO layers, structural disorder can further enhance carrier scattering, making this interplay more pronounced and potentially strengthening the oscillatory dynamic response. However, the accompanying increase in optical loss may also limit ENZ-related field-enhancement effects. A detailed analysis of how individual material parameters influence the oscillatory dynamics is provided in our previous work~\cite{Jaeik_TCO}.

\begin{figure}[htbp]
  \centering
  \includegraphics[width=1\textwidth]{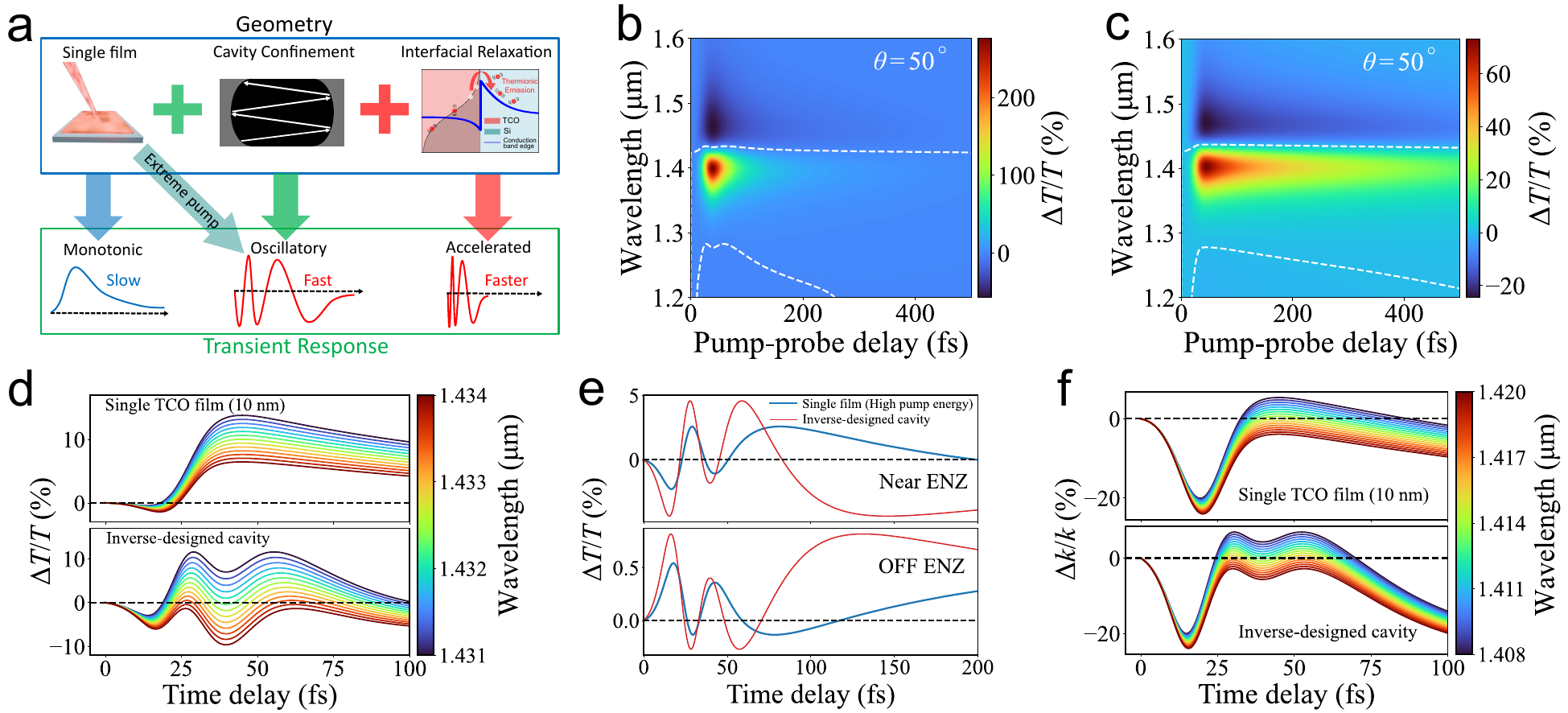}
    \caption{
    \textbf{Acceleration of oscillatory transient response enabled by interfacial relaxation control.}
    \textbf{(a)} Conceptual illustration linking cavity confinement and interfacial relaxation to distinct transient responses, from monotonic to oscillatory dynamics, and their acceleration through interfacial carrier transfer.
    \textbf{(b),(c)} Simulated transient transmittance $\Delta T/T$ as a function of pump-probe delay and wavelength at a $50^\circ$ probe incidence angle for the inverse-designed cavity and a 10-nm single TCO film, respectively. The inverse-designed cavity exhibits stronger modulation, faster recovery, and more pronounced oscillatory features at both near and off-ENZ zero-crossing regions (dotted lines).
    \textbf{(d)} Transient $\Delta T/T$ at selected probe wavelengths, showing faster oscillations and an increased number of cycles in the inverse-designed cavity compared to the single film.
    \textbf{(e)} Comparison of the transient responses of the inverse-designed cavity and a single film on fused silica under matched electron-temperature conditions, highlighting the shorter oscillation periods in the inverse-designed cavity induced by spatially nonuniform relaxation. A matched electron-temperature condition is enforced for a direct comparison of the oscillatory dynamics and is achieved by applying a higher pump fluence to the single TCO film.
    \textbf{(f)} Transient change in the imaginary part of the refractive index, $\Delta k/k$, under the excitation conditions used for the unmatched electron-temperature case in (d), revealing the close correspondence between the oscillatory transmittance and extinction-coefficient dynamics.
    }
  \label{fig:fig_5}
\end{figure}

As a result, the enhanced ENZ confinement achieved by the inverse-designed cavity can produce a far more intricate dynamic response compared to the single-film geometry, manifested as oscillatory transmittance dynamic responses. Figure~\ref{fig:fig_5}(a) provides a conceptual summary of the transient optical response explored in this section. In a single TCO film, GW/cm$^2$-level pump excitation produces a moderate electron-temperature increase and a single-peaked transient response~\cite{Kinsey_AZO_optica, Boyd_TCO_science}. Introducing cavity-enhanced confinement strengthens light absorption in the TCO layer through the ENZ-mediated amplification, allowing the transient modulation to become oscillatory~\cite{Jaeik_TCO}. While individual oscillation periods are drastically shortened compared to the single-peaked response, incorporating interfacial carrier relaxation can further shorten the durations. Together, these effects illustrate the central design principle for engineering the dynamic TCO response: combining ultrathin geometry, cavity-enhanced ENZ confinement, and interfacial relaxation can transform a single-peaked transient modulation into an oscillatory optical response with substantially shortened modulation timescales. It should be noted, however, that oscillatory responses are not exclusive to cavity geometries. In single-film TCOs, similar oscillatory dynamics, first reported experimentally in Ref.~\cite{Moti_CLEO}, can emerge under TW/cm$^2$-level or stronger pump excitation through the competing evolution of $\omega_\mathrm{p}$ and $\Gamma$~\cite{Jaeik_TCO, Pendry_inter_intra}, Auger-type excitations~\cite{Pendry_inter_intra, Pendry_Auger_LSA}, the interplay between nonlinear interband transitions and intraband dynamics~\cite{Mustafa_inter_intra, Jaffray_oscillatory}, or intraband excitation of electrons beyond the inflection point of the conduction band~\cite{Noa_oscillatory}. Nevertheless, the cavity geometry remains essential for translating these dynamics into practical on-chip platforms, where the available pump power is limited.

Transmission-dynamics maps calculated under identical pump powers, using a 60$^\circ$ ENZ pump and a 50$^\circ$ probe, show that the inverse-designed cavity exhibits stronger modulation and markedly faster relaxation (Figure~\ref{fig:fig_5}(b)) than the single TCO film (Figure~\ref{fig:fig_5}(c)). The defining feature, however, lies in the zero-crossings of $\Delta T$ (marked by dotted lines), where the signal undergoes a sign reversal. The dotted lines further highlight that these zero-crossings occur in two spectral regions, near the ENZ wavelength and in an off-ENZ range, both of which exhibit a more complex oscillatory response in the inverse-designed cavity, consistent with the physical picture illustrated in Figure~\ref{fig:fig_5}(a). Figure~\ref{fig:fig_5}(d) highlights the near-ENZ zero-crossing region for the single TCO film and the inverse-designed cavity. The colorbar denotes the probe wavelength, with the corresponding colored curves showing the transient transmittance response at each wavelength. Compared with the single film, the inverse-designed cavity exhibits stronger oscillatory modulation with a shorter temporal period, reflecting the faster dynamics enabled by cavity-enhanced ENZ confinement. This behavior goes beyond the simple intuition that stronger light confinement should merely increase the modulation amplitude at the cost of slower relaxation---a picture that remains valid for probe wavelengths away from the zero-crossing regions. Instead, the response becomes more intricate because the competition between pump-induced change in refraction and the simultaneous increase in absorption becomes more pronounced.

We examined the physical interpretation and applicability of the thermalized-electron model in detail in our previous work ~\cite{Jaeik_TCO}, which provides the foundation for the present study. Although it has recently been demonstrated that nonthermal, multiphoton~\cite{Mustafa_inter_intra, Jaffray_oscillatory}, and Auger processes~\cite{Pendry_inter_intra, Pendry_Auger_LSA} can play a role in TCO dynamics under high pump fluence, the intraband thermalized-electron response in this work provides a physically meaningful baseline. Particularly, under the ENZ pump excitation considered here ($\lambda \approx 1425$\,nm), the photon energy is approximately $0.87$\,eV, whereas the typical TCO bandgap is approximately $3.5$--$4.0$\,eV. Interband excitation therefore requires four to five-photon absorption and is expected to be substantially weaker than the intraband response under the present conditions. The estimated electron--electron scattering time is approximately $5$--$6$\,fs, leading to thermalization within approximately $10$--$20$\,fs, after a few scattering events. The thermalized-electron model should therefore capture the dominant dynamics after the initial thermalization period, although additional mechanisms may cause slight deviations.

Importantly, our simulations show that thermalized-electron dynamics alone can generate the oscillatory transmittance response, even without interband or nonthermal processes. This calculated response therefore provides a baseline against which contributions from additional mechanisms can be identified. In our previous work, comparison of this framework with experimental pump--probe measurements under strong excitation showed that the model qualitatively reproduced the measured transmission and reflection dynamics, including the unusually rapid transmission relaxation~\cite{Jaeik_TCO}. The remaining deviations at later delays may indicate additional changes in carrier density and damping, including possible Auger-type carrier-injection processes. Thus, the thermalized-electron model captures the dominant experimental trends while providing a basis for distinguishing thermalized-electron dynamics from additional high-fluence mechanisms.

Furthermore, we have earlier investigated the underlying mechanism, material dependence, and tunability of this oscillatory response in detail ~\cite{Jaeik_TCO}. The sensitivity analysis systematically varied the carrier concentration, effective mass, conduction-band nonparabolicity, peak electron temperature, and Drude-damping parameters. Increasing the carrier concentration, effective mass, or nonparabolicity shifts the oscillatory region spectrally, while the damping parameters further control its spectral and temporal characteristics. Despite these shifts, the oscillatory response persists over a broad parameter range, showing that the behavior in Figure~\ref{fig:fig_5} is robust and not restricted to a specific set of material parameters.

To distinguish the effects of enhanced electron heating from those of interface-controlled relaxation, Figure~\ref{fig:fig_5}(e) compares the transmittance dynamics of the inverse-designed cavity and a 10\,nm single TCO film on fused silica under conditions of matched peak electron temperature. While ENZ confinement introduces a richer, more complex transient response, interface-controlled relaxation has a pronounced effect on the duration of oscillation periods. The single-film response is calculated using a pump pulse twice as strong as that of the cavity to achieve the same electron temperature rise. Under this peak temperature-matched condition, both structures show five modulation periods, defined as the 0--peak--0 interval between successive zero-crossings of $\Delta T$. This allows for a straightforward comparison of their oscillation timescales. The first two modulation periods occur during the electron temperature rise and are therefore mostly determined by the pump-pulse duration. In contrast, the durations of later oscillation periods are governed by the electron temperature relaxation dynamics. Because interfacial carrier transport accelerates relaxation, the successive modulation periods in the inverse-designed cavity are shortened by $1.3$--$4\times$ relative to those of the single TCO film, enabling substantially faster oscillatory modulation. Table~\ref{tab:mod_periods} summarizes the duration of each modulation period for the two geometries. Notably, the first two modulation intervals have comparable durations between geometries because they are primarily governed by the pump pulse. Subsequent modulation intervals of the inverse-designed cavity are shorter than those of the single film, with the fourth and fifth showing the largest relative compression.

\begin{table}[H]
\centering
\caption{Enhanced modulation period durations in the inverse-designed TCO cavity at near-ENZ probe (units: fs).}
\label{tab:mod_periods}
\begin{tabular}{lccccc}
\hline
Geometry & Period 1 & Period 2 & Period 3 & Period 4 & Period 5 \\
\hline
Inverse-designed cavity & 21.6 & 13.2 & 10.4 & 37.6 & 495 \\
Single TCO film         & 22.9 & 13.8 & 13.9 & 149.1 & 944.5 \\
\hline
\end{tabular}
\end{table}

Overall, by combining strong ENZ confinement with interface-controlled relaxation, the inverse-designed cavity reaches the high electron temperatures needed for oscillatory modulation while also compressing the modulation periods to support much faster dynamics. Finally, Figure~\ref{fig:fig_5}(f) shows that the oscillatory transmittance response is closely associated with the transient evolution of the imaginary refractive index, $k$. For both the single film and the inverse-designed cavity, the modulation of $k$ follows trends similar to the transmittance dynamics shown in Figure~\ref{fig:fig_5}(d), indicating that changes in optical loss strongly govern the dynamic response. The ability to modulate the intrinsic optical properties over multiple periods underscores the promise of inverse-designed TCO cavities as time-varying photonic media, while revealing the critical role of interface engineering in tailoring the material response time. In particular, their potential capability to support sequential non-adiabatic events, including time refraction and time reflection, could open an important direction for future investigation. Temporally and spectrally co-optimized cavity structures, therefore, provide a practical route toward dynamic photonic functionalities that have largely remained theoretical.

\section{Discussion}\label{sec3}

The theoretical analysis presented here establishes inverse-designed TCO cavities as a promising platform for achieving ultrafast, large-amplitude optical modulation beyond that of single TCO films and conventionally designed cavities, through the combined control of ENZ field confinement and interfacial carrier relaxation. Unlike conventional cavity geometries, where enhanced absorptance does not necessarily translate into strong local electron heating in TCOs, our inverse-designed structures concentrate optical energy directly within an ultrathin TCO layer while preserving an efficient pathway for carrier extraction into the adjacent Si. We theoretically show that this combination produces a dynamic response that is both stronger and faster than that of single TCO films and intuitively designed TPP-based cavities. Importantly, the response is not governed by optical confinement alone, but emerges from the co-design of field localization, TCO thickness, and interfacial properties.

The structural disorder-aware correction framework highlights an important practical requirement for translating the theoretically optimized inverse-designed TCO cavities into experimentally feasible devices. The optical properties of ultrathin doped oxides can deviate strongly from thickness-invariant optical constants, particularly near disordered interfaces, causing designs based on bulk-like or thick-film parameters to miss the intended resonance after fabrication. The proposed GDL-based re-optimization loop compensates for these deviations by reconstructing the effective optical constants of the realistic material system and re-optimizing the dielectric stack to restore the target resonance. Although this study focuses on numerically evaluating the optical response of the theoretical inverse-designed cavities, the proposed correction strategy provides a practical pathway toward the experimental realization of such cavities.

A central implication of this work is that the interfaces should be treated as active design elements in time-varying photonic structures. In ultrathin TCO layers, the electron temperature dynamics are strongly influenced by carrier transport across nearby boundaries, such that the surrounding material environment can either accelerate relaxation or introduce long-lived background tails. The interface analysis shows that optimizing interfacial barriers can preserve larger peak electron temperatures while maintaining fast recovery. In addition, we theoretically demonstrate that heavy doping of the adjacent Si layer can suppress slow relaxation tails by promoting field-emission-like transport. These results suggest that interface engineering through deposition conditions, annealing, and doping profiles offers practical routes to tailor not only the static optical resonance but also the temporal response.

Furthermore, the oscillatory transient transmittance predicted in these structures suggests a pathway toward time-varying photonic media operating in the non-adiabatic regime. Extreme electron heating in the TCO layer of the inverse-designed cavity induces concurrent modulation of the plasma frequency and Drude damping, resulting in sign-changing transmittance dynamics over multiple modulation intervals. Interfacial carrier transfer from the TCO to Si accelerates the relaxation dynamics, further shortening the oscillatory modulation intervals. These shortened modulation dynamics are particularly relevant for non-adiabatic photonic phenomena, including time refraction, time reflection, frequency conversion, and photonic-time-crystal-like responses, where material-property modulation on the optical-cycle timescale is essential.

The present predictions, however, should be interpreted as a qualitative design baseline for the thermalized-electron response. Under strong experimental excitation, additional multiphoton interband or Auger-type carrier processes may produce deviations from the calculated dynamics. This baseline can therefore help distinguish thermalized-electron dynamics from contributions of additional high-fluence mechanisms, as illustrated in our previous work~\cite{Jaeik_TCO}. Overall, this work establishes inverse-designed, interface-engineered TCO cavities as a promising platform for dynamic photonics, including ultrafast modulators, optical switches, and non-adiabatic photonic systems, while providing design principles that can be extended to hybrid material systems.

\section*{Methods}
\subsection*{Model description}

The coupled evolution of electron temperature, interfacial carrier transport, and the resulting modification of the TCO permittivity determine the transient optical response of the multilayer cavity. In the inverse-designed structures considered here, the nonlinear response is dominated by the TCO layer. The modeling framework integrates the key physical processes governing the transient response. A spatially resolved two-temperature model first determines the spatiotemporal electron-temperature evolution, while interfacial carrier extraction is self-consistently accounted for through a thermionic boundary condition. Then, the temperature-dependent permittivity of the TCO is calculated from the evolution of the electron temperature. Finally, the transient optical modulation is estimated.

\subsection*{TMM-based inverse-design algorithm}

The differentiable TMM solver and gradient-based optimization routine were implemented using custom Python code. We fixed the material sequence and the 10\,nm TCO thickness, while optimizing the thicknesses of the five Si/SiO$_2$ layers and the TCO carrier concentration by minimizing the mean-squared error between the calculated and target absorptance spectra. The optical model assumes planar layers with semi-infinite incident and substrate media and enforces continuity of the tangential electric and magnetic fields at each interface. The optimization was continued until the mean-squared-error loss converged.

\subsection*{Two-temperature model}

The electron and lattice temperature evolution over time are described using the coupled two-temperature equations,
\begin{equation}
C_e(T_e)\frac{\partial T_e}{\partial t}
=
\frac{\partial}{\partial x}
\left(
\kappa(T_e)\frac{\partial T_e}{\partial x}
\right)
-
g_{ep}(T_e)(T_e-T_\mathrm{L})
+
H(x),
\label{eq:TTM_e}
\end{equation}
\begin{equation}
C_\mathrm{L}(T_\mathrm{L})\frac{\partial T_\mathrm{L}}{\partial t}
=
g_{ep}(T_e)(T_e-T_\mathrm{L}),
\label{eq:TTM_L}
\end{equation}
where $T_e$ and $T_L$ are the electron and lattice temperatures, $C_e$ and $C_\mathrm{L}$ are the electron and lattice heat capacities, $\kappa$ is the electronic thermal conductivity, $g_{ep}$ is the electron-phonon coupling coefficient, and $H(x)$ is the absorbed optical power density.

The spatially resolved two-temperature equations were solved using a modified implementation of the Python-based NTMpy package~\cite{NTMpy}. We set the initial electron and lattice temperatures to 300\,K and imposed zero-heat-flux Neumann boundary conditions at the external boundaries. At ordinary material interfaces, continuity of temperature and normal heat flux was enforced. At the TCO/Si interface, the electron heat-flux continuity condition was replaced by the thermionic-emission boundary condition described later, while the conventional continuity condition was retained for the lattice temperature. We selected the integration time step according to the NTMpy stability criterion, $\Delta t < 2/|\lambda|_{\max}$, where $|\lambda|_{\max}$ is the largest-magnitude eigenvalue of the discretized thermal-diffusion and coupling matrix. To ensure stability across the full temperature range, we evaluated this criterion for each layer and used the smallest permitted time step among all layers.

The heat source $H(x)$ is obtained from spatially resolved transfer-matrix calculations of the multilayer structure. The temporal Gaussian profile of the pump pulse is Fourier transformed into the frequency domain, enabling calculation of the spatially distributed absorbed energy for each frequency component. This yields a spatially resolved heat source that captures the non-uniform energy deposition induced by the ultrafast pump pulse.

The temperature dependence of the electronic heat capacity is obtained from
\begin{equation}
C_e(T_e) =
\int_{0}^{\infty}
\frac{\partial f_\mathrm{FD}(\mu(T_e),T_e)}{\partial T_e}
\, g(E)\, E \, \de E,
\end{equation}
where $f_\mathrm{FD}$ is the Fermi-Dirac distribution and $g(E)$ is the electronic density of states. The derivative of $f_\mathrm{FD}$ with respect to temperature acts as a weighting function, selecting the electrons that strongly respond to thermal excitation. The equation, therefore, quantifies how these thermally active electrons redistribute in energy as electron temperature evolves and how much energy this redistribution carries. The lattice heat capacity is taken as a constant, \( C_\mathrm{L} = 2.54 \times 10^6 \, \text{J m}^{-3} \text{K}^{-1} \), since the lattice temperature varies only weakly over the timescale of interest.

The electron-phonon coupling coefficient is written as
\begin{equation}
g_{ep} =
\frac{\pi \kB}{\hbar\, g(E_\mathrm{F})}
\lambda_{ep}\langle(\hbar\omega_{ph})^2\rangle
\int
g^2(E)
\left(-\frac{\partial f_\mathrm{FD}}{\partial E}\right)\de E,
\end{equation}
where $\kB$ is the Boltzmann constant, $\hbar$ is the reduced Planck constant, and $E_\mathrm{F}$ is the Fermi energy.

The parameter $\lambda_{ep}$ is the electron-phonon mass enhancement factor, characterizing the strength of electron-phonon interactions, and $\langle(\hbar\omega_{ph})^2\rangle$ is the second moment of the phonon spectrum, representing the average squared phonon energy weighted by the phonon density of states. Their product, $\lambda_{ep}\langle(\hbar\omega_{ph})^2\rangle$, is taken as $5.25 \times 10^{-4}\,\mathrm{eV}^2$ following Ref.~\cite{Boyd_TCO_science}. The integral term emphasizes that electron-phonon coupling is dominated by electronic states near the Fermi level, where the derivative of the Fermi-Dirac distribution is largest, and therefore captures the effective coupling strength of thermally active carriers.

\subsection*{Thermionic emission boundary condition}

Importantly, the interfacial boundary condition is essential in the present two-temperature framework, as energy transport across interfaces plays a dominant role in the relaxation dynamics.
We explicitly treat the carrier transport across the TCO/Si interface through the thermionic emission mediated electron heat flux at the interface,
\begin{equation}
\kappa \left. \frac{\partial T_e}{\partial x} \right|_{x = x_{\mathrm{interface}}} = - (E_\mathrm{F} + \phi)\left. \frac{\de J_{\text{esc}}}{\de t} \right|_{x = x_{\mathrm{interface}}},
\label{eq:Interfacial_BC}
\end{equation}
where $\kappa$ is the electronic thermal conductivity, $E_\mathrm{F}$ is the Fermi energy, $\phi$ is the interfacial barrier energy of TCO/Si interface, estimated to be 0.45\,eV, $\de J_{\text{esc}}/\de t$ is the thermionic emission rate of carriers crossing the barrier, and $J_{\text{esc}} = N_{\text{esc}}/(\pi R_1 R_2)$ is the emitted electron number per unit area, where $R_1$ and $R_2$ are the semi-major and semi-minor axes of the elliptical emission region, respectively.

In the absence of space-charge effects, thermionic emission follows the classical Richardson-Dushman relation,
\begin{equation}
\frac{\de J}{\de t} =
\frac{A_0}{e} T_e^2
\exp\left(-\frac{\phi}{\kB T_e}\right),
\end{equation}
where $A_0$ is the Richardson constant dependent on the material~\cite{Thermionic_Sim}, $e$ is the elementary charge, and $\kB$ is the Boltzmann constant. This expression assumes that the chemical potential remains close to the Fermi energy and that the emitted carriers do not significantly perturb the electrostatic environment near the interface.

Under strong optical excitation, however, the electron temperature rises substantially, resulting in a large population of emitted carriers. In this regime, accurate modeling requires accounting for both the electron temperature-dependent chemical potential, $\mu(T_e)$, and the space-charge potential that builds up during emission. These effects are especially important in TCOs, where high electron temperatures produce substantial chemical-potential shifts and pronounced thermal broadening of the electron energy distribution. The thermionic emission rate is therefore modified as
\begin{equation}
\frac{\de J_{\text{esc}}}{\de t}
=
A (k_{\mathrm B} T_{\mathrm e})^2
\exp\left[
-\frac{E_\mathrm{F} - \mu(T_{\mathrm e}) + \phi + \phi_{\mathrm{sc}}}{k_{\mathrm B} T_{\mathrm e}}
\right],
\label{eq:N_rate}
\end{equation}
where $E_\mathrm{F}$ is the Fermi energy and $\mu(T_e)$ is the electron temperature-dependent chemical potential. The prefactor $A=4\pi m^\ast h^{-3}$ depends on the material through the effective electron mass $m^\ast$, and $h$ is Planck's constant. The additional term $\phi_{sc}$ represents the electrostatic potential energy arising from the accumulated charge of emitted electrons.

The space-charge potential energy is estimated (in Gaussian units) as
\begin{equation}
\phi_{sc} \approx \frac{a\, N_{\text{esc}}\, e^2}{R_1},
\end{equation}
where $N_{\mathrm{esc}}$ is the total number of emitted electrons, and $a$ is a geometrical factor, taken here as $a=16/(3\pi\ve_r)$ for a uniformly emitting region, with $\ve_r$ denoting the static relative permittivity of the emission region.

Assuming the square-wave temporal profile of the electron temperature during the emission for a pump pulse duration $\tau$, Eq.~\ref{eq:N_rate} can be integrated analytically to obtain the total emitted number of electrons~\cite{Thermionic_space_charge},
\begin{equation}
N_{\text{esc}} = \frac{\kB T_e}{ae^2/R_1} \log \left[ 1 + A \tau \pi R_2 ae^2 \kB T_e \exp\left( \frac{-E_\mathrm{F} + \mu(T_e) - \phi}{\kB T_e} \right) \right],
\label{eq:N_yield}
\end{equation}

The final expression accounts for the increase in the effective emission barrier caused by space-charge buildup and the electron-temperature-dependent chemical-potential shift. In the present work, we simplify the emission geometry by taking $R_1 = R_2$, corresponding to a circular emission region in the interface plane. This expression shows that the total yield is determined by the competition between thermionic emission and space-charge buildup, which progressively raises the effective emission barrier and suppresses further carrier extraction.

This formulation provides a physically consistent description of thermionic emission in TCO/Si junctions, accounting for both space-charge buildup and temperature-induced changes in electronic properties under strong excitation. Through the interfacial boundary condition in Eq.~\ref{eq:Interfacial_BC}, the extracted carrier flux is coupled to the interfacial electron-temperature drop, enabling the effect of interfacial carrier transport on the electron-temperature dynamics to be self-consistently evaluated.

\subsection*{Electron temperature–dependent permittivity of TCO}

In the absence of optical modulation, the complex permittivity of the TCO is described by the Drude model, with a carrier-density-dependent plasma frequency:
\begin{equation}
\varepsilon(\omega,n_e)
=
\varepsilon_{\infty}
-
\frac{\omega_p^2(n_e)}{\omega^2+i\Gamma\omega},
\qquad
\omega_p(n_e)
=
\sqrt{\frac{n_e e^2}{\varepsilon_0 m^*}}.
\label{eq:Drude}
\end{equation}

Here, $n_e$ is the carrier concentration, $e$ is the elementary charge, $\varepsilon_0$ is the vacuum permittivity, and $m^*$ is the effective electron mass. The carrier-density-dependent plasma frequency, $\omega_p(n_e)$, captures the strength of the free-carrier response, whereas the Drude damping rate, $\Gamma$, accounts for carrier-scattering loss at room temperature. The values of $\Gamma$ and the background permittivity, $\varepsilon_{\infty}$, are obtained from measured AZO films.

To describe the transient permittivity modulation of the TCO layer under strong excitation, we model its complex permittivity using a temperature-dependent Drude formalism. The key quantities in this model are the plasma frequency and Drude damping rate, both of which evolve with electron temperature due to carrier redistribution in the non-parabolic conduction band and enhanced electron scattering.

The non-parabolic conduction-band dispersion is written as
\begin{equation}
\frac{\hbar^2 k^2}{2m^*} = E + C E^2,
\end{equation}
where $E$ is the electron energy, $k$ is the wavevector, $m^*$ is the effective mass at the conduction-band minimum, and $C$ is the non-parabolicity parameter. In the present model, we use $m^* = 0.24m_e$ and $C = 0.27~\mathrm{eV}^{-1}$, where $m_e$ is the electron mass.

These values represent a reference parameter set for TCOs based on previously reported material parameters~\cite{Boyd_TCO_science}. Because the effective mass and nonparabolicity can vary with TCO composition and carrier concentration, their influence was systematically examined in our previous work~\cite{Jaeik_TCO}. That sensitivity analysis varied the carrier concentration, effective mass, nonparabolicity, peak electron temperature, and temperature dependence of the Drude damping over broad ranges. The oscillatory response persisted throughout these parameter variations, while its spectral position and temporal characteristics shifted systematically. The parameters adopted here therefore define the specific operating wavelengths and timescales, whereas the underlying oscillatory behavior is robust across a broad material parameter space.

From this non-parabolic dispersion, the electronic density of states is obtained as
\begin{equation}
g(E) =
\frac{\sqrt{2}}{\pi^2}
\left(\frac{m^*}{\hbar^2}\right)^{3/2}
(E+CE^2)^{1/2}(1+2CE).
\end{equation}

The electron temperature-dependent permittivity of TCO is then written in Drude form as
\begin{equation}
\ve(\omega,T_e) =
\ve_{\infty}
-
\frac{\omega_p^2(T_e)}{\omega^2 + \ii\Gamma(T_e)\omega},
\label{eq:Drude_Te}
\end{equation}
where $\ve_{\infty}$ is the high-frequency permittivity, $\omega$ is the angular frequency of the electromagnetic field, $\omega_p(T_e)$ is the temperature-dependent plasma frequency, and $\Gamma(T_e)$ is the temperature-dependent Drude damping.

For a non-parabolic conduction band, the plasma frequency cannot be written in the simple form $\omega_p=\sqrt{n_e e^2/(\varepsilon_0 m^*)}$ because the carrier effective mass is not constant across the occupied states. Instead, $\omega_p(T_e)$ is derived from the linearized Boltzmann transport equation to be~\cite{Boyd_TCO_science, Chang_Nature_photonics, Boyd_TCO_Nature_photonics}
\begin{equation}
\omega_p^2(T_e) =
\frac{-e^2}{3 m^\ast \varepsilon_0 \pi^2}
\int_{0}^{\infty}
\frac{\de E}{1 + 2 C E}
\left(
\frac{2m^*}{\hbar^2}  (E + C E^2)
\right)^{3/2}
\frac{\partial f_\mathrm{FD}(E,\mu(T_e),T_e)}{\partial E}
,
\end{equation}
where $e$ is the elementary charge, $\varepsilon_0$ is the vacuum permittivity, and $f_\mathrm{FD}$ is the Fermi-Dirac distribution function. Physically, this expression shows that the plasma frequency is governed not only by the total carrier density but also by the thermally active carriers near the chemical potential. In a non-parabolic conduction band, these carriers have an energy-dependent effective mass, which modifies the free-carrier response. The factor $-\partial f_{\mathrm{FD}}/\partial E$ acts as a weighting function centered around the chemical potential, selecting the electronic states that contribute most strongly to the thermally induced change in plasma frequency.

The chemical potential entering the Fermi-Dirac distribution function is obtained by self-consistently enforcing carrier-density conservation during intraband excitation of TCO,
\begin{equation}
n_e =
\int_{0}^{\infty}
f_\mathrm{FD}(E,\mu(T_e),T_e)\, g(E)\, \de E,
\end{equation}
where $n_e$ is the total electron density.

The second temperature-dependent quantity in Eq.~\ref{eq:Drude_Te} is the Drude damping rate, which describes carrier-scattering losses arising from phonons, impurities, grain boundaries, surface roughness, and microstructural disorder,
\begin{equation}
\Gamma(T_\mathrm{e}) = \Gamma_{\mathrm{RT}} \left[ 1 + \left(\frac{T_\mathrm{e}}{T_0}\right)^m \right],
\label{eq:damping_update}
\end{equation}
where $\Gamma_{\mathrm{RT}}$ is the measured Drude damping at room temperature, $m$ describes the power-law dependence of the damping on $T_\mathrm{e}$, and $T_0$ is a fitting parameter that sets the characteristic temperature scale of this dependence. Because both $m$ and $T_0$ depend sensitively on the sample-specific microstructure, their values must be calibrated against experimental measurements. In the present model, we set $m=2$ and use $T_0=4000$\,K, obtained by fitting the experimental data reported in Ref.~\cite{Moti_TCO}. A detailed description of the simulation-experiment comparison is provided in our previous work~\cite{Jaeik_TCO}.

In contrast to noble metals, where the temperature dependence of the damping often remains modest, TCOs can reach electron temperatures comparable to or exceeding the Fermi temperature under strong pumping, making the damping enhancement a major contributor to the transient optical response~\cite{Sivan_damping, TCO_scattering_Nanophotonics}.

\subsection*{Transient transmittance modulation}

Finally, the transient optical modulation is evaluated by incorporating the temperature-dependent TCO permittivity into the cavity optical response. Because the cavity nonlinearity is dominated by the TCO layer, the resulting transmittance dynamics are determined primarily by the temporal evolution of the TCO complex permittivity.

The dependence of the transmittance $T$ on the complex permittivity $\varepsilon=\varepsilon'+i\varepsilon''$ can be linearized through a first-order expansion:
\begin{equation}
\delta \ln T =
\frac{\partial \ln T}{\partial \varepsilon'} \delta \varepsilon'
+
\frac{\partial \ln T}{\partial \varepsilon''} \delta \varepsilon'',
\label{eq:d_transmission}
\end{equation}

Assuming that the derivative terms remain approximately constant over the transient evolution, this expression can then be integrated to obtain the time-dependent transmittance modulation
~\cite{Three_temp_review, Ncomm_del_T_Eq, TPP_modeling}
\begin{equation}
\ln\left(\frac{T_\mathrm{mod}}{T}\right)
=
\frac{\partial \ln T}{\partial \varepsilon'} \Delta \varepsilon'(T_e)
+
\frac{\partial \ln T}{\partial \varepsilon''} \Delta \varepsilon''(T_e).
\label{eq:transmission}
\end{equation}
In this framework, the derivatives $\partial \ln T / \partial \varepsilon'$ and $\partial \ln T / \partial \varepsilon''$ represent the cavity sensitivities to perturbations in the real and imaginary components of the TCO permittivity, respectively~\cite{Jaeik_TCO}.

The first-order approximation is well suited to the present system, where the cavity resonance remains sufficiently broad. The Supplementary Information of our previous work~\cite{Jaeik_TCO} further supports its applicability under strong excitation, where the same temperature-dependent optical framework qualitatively reproduced experimentally measured transmission and reflection dynamics. We made this comparison in a regime where the electron temperature approaches the Fermi temperature and is therefore comparable to the conditions considered here. For much higher-$Q$ cavities, however, even small permittivity changes can shift the resonance substantially, leading to time-dependent variations in the cavity sensitivity and invalidating the assumption used to obtain Eq.~\ref{eq:transmission} from Eq.~\ref{eq:d_transmission}. In such systems, a more rigorous approach based on direct time-domain solutions of Maxwell's equations self-consistently coupled to electron-temperature-dependent nonlinear source terms is needed~\cite{LEDD_rigorous, TTM_FDTD}. Such rigorous formulations would allow non-adiabatic effects such as time refraction, time reflection, and frequency conversion to be directly calculated in space-time photonic systems, making them an important direction for future study. Beyond engineering these non-adiabatic processes within TCOs, alternative material strategies could offer a complementary route for further improving the ultrafast response. MXenes, in particular, support rapid nonthermal relaxation through direct coupling of nonthermal electrons to coherent $A_{1g}$ phonons~\cite{MXene_A1g, Science_MXene}. Applying the present inverse-design framework to MXene/TCO hybrids could combine this rapid relaxation pathway with the strong ENZ response of TCOs, providing a promising route toward faster and stronger modulation~\cite{Jaeik_MXene}.

\section*{Code and Data Availability}
The code and data supporting the findings of this study are publicly available on Figshare at https://doi.org/10.6084/m9.figshare.34695303.

\section*{Acknowledgement}
This material is based upon work supported by the U.S. Department of Energy, Office of Science, Office of Basic Energy Sciences, Division of Materials Sciences and Engineering under Award Number DE-SC0017717 (transparent conducting oxides characterization and optical properties) and a collaborative project with DEVCOM Army Research Lab “Ultrafast Space-Time Photonics and Single-Photon Optical Modulators,” and the U.S. Army Research Office through the Department of Defense Multidisciplinary University Research Initiative (MURI), “N-Dimensional (ND) Space-Time Metamaterials (ND-STMM)” under Award No. W911NF-26-1-A258. The authors thank Ian Hoffman for his valuable assistance with the optical characterization of the TCO films.

\section*{Competing Interests}
The authors declare no competing interests.

\nocite{
efimov1979analytical,
brendel1992infrared,
peiponen1991kramers,
orosco2018optical,
nordebo2024time,
franta2018universal,
stenzel2019beta,
faddeeva1961tables}

\bibliographystyle{naturemag}
\bibliography{reference}

\end{document}